\documentclass{aastex63}
\usepackage{graphicx}
\usepackage{graphics}
\usepackage{dcolumn}
\usepackage{bm}
\usepackage{hyperref}
\usepackage{comment}
\usepackage{amsmath}

\shorttitle{Habitable zones around almost extremely spinning black holes}
\shortauthors{Bakala et al.}

\begin{document}

\title{Habitable zones around almost extremely spinning black holes (black sun revisited)}

\correspondingauthor{Pavel Bakala}
\email{pavel.bakala@physics.cz, pbakala@hvezdaren.org}

\author[0000-0003-0951-8597]{Pavel Bakala}
\affiliation{Research Centre for Computational Physics and Data Processing, Faculty of Philosophy \& Science \\
Bezru\v{c}ovo n\'am.~13, CZ-746\,01 \\
Opava, Czech Republic}
\affiliation{M. R. \v{S}tef\'anik Observatory and Planetarium\\
Sl\'adkovi\v{c}ova 41, 920 01 Hlohovec\\
Slovak Republic}
\affiliation{INAF -  Osservatorio Astronomico di Roma\\
Via Frascati, 33, Monteporzio Catone, 00078  Roma
\\Italy}

\author{Jan Do\v{c}ekal}
\affiliation{Research Centre for Computational Physics and Data Processing, Faculty of Philosophy \& Science \\
Bezru\v{c}ovo n\'am.~13, CZ-746\,01 \\
Opava, Czech Republic}

\author{Zuzana Turo\v{n}ov\'a}
\affiliation{Research Centre for Computational Physics and Data Processing, Faculty of Philosophy \& Science \\
Bezru\v{c}ovo n\'am.~13, CZ-746\,01 \\
Opava, Czech Republic}

\begin{abstract}
We analyzed the thermodynamics of hypothetical exoplanets at very low Keplerian circular orbits in close vicinity of rapidly spinning supermassive black holes. Such black hole exoplanets are heated by strongly blueshifted and focused flux of the incoming cosmic microwave background (CMB) and cooled by the cold part of the local sky containing the black hole shadow. This gives rise to a temperature difference, which can drive processes far from thermodynamic equilibrium in a hypothetical life form inhabiting black hole exoplanets, similar to the case of a planet heated by the radiation of the parent star and cooled by the night sky.
We found that for a narrow range of radii of very low Keplerian circular orbits and for very high spin of a supermassive black hole, the temperature regime of the black hole exoplanets corresponds to the habitable zone around standard stars. The thermodynamics of black hole exoplanets therefore, in principle, does not exclude the existence of life based on known biology. The peak of the multiblackbody spectral profile of the CMB heating the exoplanet is located in the ultraviolet band, but a significant fraction of the flux comes also in the visible and infrared bands. The minimum mass of a black hole ensuring the resistance to tidal disruption of an Earth-like exoplanet orbiting in the habitable zone is estimated to $1.63 \cdot 10^8 \, m_{\odot}$.
\end{abstract}


\section{Introduction}

Exoplanets and the related possibility of extraterrestrial life are certainly one of the hottest topics of contemporary astronomy, astrophysics, and astrobiology. Naturally, the basic condition for the existence of imaginable biological forms is a sufficient supply of energy, and therefore a reasonable surface temperature of the exoplanet. Apart from biochemical details, the sufficient temperature of an exoplanet is not enough. Living organisms decrease their entropy. On the contrary, the entropy of the surrounding universe increases by waste heat of their biological processes, which are driven by the energy of the incoming radiation \citep{Schrodinger1944-SCHWIL-5}. The existence of a temperature gradient between a low-entropy radiation source that heats the (exo)planet and a cold part of the sky that allows the discharge of waste heat is a necessary condition for the existence of life processes. We can also consider alternative scenarios of extraterrestrial life (see, e.g., \cite{2017arXiv170503394S,2018PDU....22...74H}), but this will not affect the essence of our thermodynamic considerations. Such a situation corresponds very well to the case of an (exo)planet heated by a standard parent star.

However, a more exotic scenario is discussed and analyzed by \cite{BlackSun}. As a source of the energy for life processes, the cosmic microwave background (CMB) is considered. Let us keep in mind, that CMB almost isotropically fills the universe, and its energy spectrum with a very high accuracy corresponds to the blackbody radiation with a temperature $T_{cmb}=2.7260\pm 0.0013$ K measured in the frame comoving with galaxies dragged by the universe expansion \citep{2009ApJ...707..916F,IsoStatCMB}. Of course, the CMB is completely irrelevant for heating planets orbiting around standard stars, and the night sky is a powerful heat sink. However, an extreme gravitational field in the vicinity of rotating black holes could radically change the situation. An exoplanet orbiting a black hole at a very low circular orbit will be heated by the strongly blueshifted CMB in the extreme gravitational field, and will be cooled by a black hole shadow with a temperature close to absolute zero. Fully relativistic numerical modeling of the CMB flux incoming from the exoplanetary local sky by \cite{BlackSun} shows that the key factor for the possibility of significant heating the exoplanet by the CMB is the very high black hole spin that shifts the position of the innermost stable circular orbit (ISCO) very close to the event horizon. The gravitational blueshift of the radiation coming from a distant universe to the exoplanetary local sky at a very low circular orbit around a rapidly spinning black hole is extremely high. Moreover, the orbital velocity at such orbits reaches a substantial fraction of the speed of light, and the incoming CMB is therefore also significantly focused and amplified by the optical effects of special relativity.

In \citep{BlackSun}, two extreme cases were subjects of the detailed numerical analysis. The first one was explicitly inspired by the science fiction film \emph{Insterstellar}, in which heroes visit the Miller's exoplanet in a close vicinity of Gargantua, a supermassive extremely fast rotating black hole with the spin $1-1.3 \cdot 10^{-14}$ \citep{thorne2014science}. According to the film plot, the time dilation factor on the exoplanetary surface reaches the value of $\sim\,61,000$, which corresponds to the position of the Keplerian circular orbit at $r = 1.0000379 \, GM/c^{2}$, i.e. very close to the ISCO and hence to the event horizon. At such an orbit, more than $99\%$ of incoming CMB flux comes from a small spot in the exoplanet local sky with the size of a few angular seconds, where the CMB frequency shift factor reaches the value of $g=275,000$. (See Figs. 3 and 4 in \citep{BlackSun}). On the contrary, the black hole shadow covers more than $ 40 \% $ of the local sky. Therefore, the heating regime is very similar to the regime of the planet heated by a standard star, when almost all of the incoming energy can be converted to useful work and thus drive the life processes (See Section III in \citep{BlackSun}, subsection Regime ``Small heating area'' for details). The density of the CMB flux coming from an extremely radiating small spot will therefore play a role analogous to the solar constant. In the analyzed case, however, the energy flux density of the CMB reaches the value of $ \Phi \approx 420 \, kW / m^{2} $.  It is approximately three hundred times more than the value of the solar constant at the orbit of the Earth. Approximating the exoplanet as a black body, we find that its estimated temperature will reach nearly 900 degrees Celsius.

The second extreme case analyzed by \cite{BlackSun}, an exoplanet at the lowest possible orbit ($ r = 6 \, GM/c^{2}$) around a non-rotating Schwarzschild black hole, displays a completely different picture (see also \cite{2014PhRvD..90l4024W}). The the black hole shadow covers $ 12.2 \% $ of the local sky, the maximum value of the frequency shift of radiation coming from the distant universe is only $ \approx 2.12$, and we cannot speak of significant angular focusing of the CMB flux incoming from the local sky. The resulting CMB flux is only about an order of magnitude larger than the CMB flux from the the sky unaffected by the gravitational field, and it is totally negligible for the heating of the planet.

The aforementioned results show that the exoplanets orbiting around the rotating black holes can be heated by a CMB amplified in a strong gravitational field, and their resulting temperature can be varied within a relatively wide range whose upper limit exceeds 1000 degrees Celsius. In this paper, we extended the numerical analysis of the CMB flux from the local sky of black hole exoplanets to identify the range of the values of the exoplanetary orbital radius and the black hole spin corresponding to the thermodynamic conditions in the habitable zone around normal stars. We approximate the hot and the cold edge of the habitable zone by the value of solar constant on the orbits of Venus and Mars, respectively \citep{HabitableZone}.

The remainder of this paper is structured as follows:  In the second section, we derive an analytical formula for the flux density of the incoming CMB as a function of the coordinates in the local sky of the exoplanet circularly orbiting the rotating black hole. Using this, we formulate the relation for the estimated temperature of the exoplanet by integrating the flux of the incoming CMB. The third section describes the numerical methods used to calculate the CMB flux, the temperature of the exoplanet and the resulting spectral profile . In the fourth section, we identify the relevant values of the radius of the exoplanet orbit and the black hole spin. The fifth section discusses the resistance of exoplanets in habitable zone to tidal disruption. In the sixth section, we analyze the resulting spectral profile of the incoming CMB. The seventh section is devoted to the analysis of relevan time scales. The final section discusses and summarizes the results.

\section{Heating a black hole exoplanet by amplified CMB}

Let us consider an exoplanet heated by a radiation source with a small angular size. The power, $W$, which transmits radiation with flux density, $\Phi$, to an exoplanet with the radius, $\mathrm{R}$, reads
\begin{equation}
 \label{eq:power}
W=\pi\,\mathrm{R}^2\Phi.
\end{equation}
To estimate the exoplanetary temperature $T$, we will further approximate the exoplanet as a black body in thermodynamic equilibrium that radiates over its entire surface.
Using the Stefan-Boltzmann law, we can write the radiated power in the form
\begin{equation}
 \label{eq:SBL}
W=\sigma\, 4\pi\,\mathrm{R}^2\, T^4\,,
   \end{equation}
where $\sigma$ is the Stefan-Boltzmann constant, $\sigma \approx 5.68\times 10^{-8}$\,W\,m$^{-2}$\,K$^{-4}$ .
Then we can obtain the relation for the estimated temperature of the exoplanet in the form
\begin{equation}
 \label{eq:exotemp}
   T=\sqrt[4]{\frac{\Phi}{4\sigma}}\,.           
	 \end{equation} 

In a local sky of an exoplanet orbiting deep in the gravitational field of a black hole, the CMB energy spectrum will be strongly influenced by the frequency shift factor $g$, which will be the result of the presence of a strong axially symmetrical gravitational field and relativistic orbital motion, and it will necessarily depends on coordinates in the local sky.
The CMB frequency shift factor in the exoplanet local sky reads
\begin{equation}
 \label{eq:factor}
	g=\frac{E_{mathrm{obs}}}{E_{0}}=\frac{\nu_{mathrm{obs}}}{\nu_{0}}\,,
   \end{equation}
where $E_{\mathrm{obs}}$ and $\nu_{mathrm{obs}}$ are the CMB photon energies and frequencies measured in the local frame associated with the exoplanet, $ E_{0} $ and $ \nu_ {0} $ are the CMB photon energies and frequencies measured in a distant universe where the CMB is (almost) isotropic.
The source intensity divided by the third power of the frequency is conserved as the Lorentz invariant \cite{DopplerIntensity}. Therefore, the locally measured spectral intensity of a photon beam of the CMB $ I_{obs} $, which was emitted in a distant universe with a spectral intensity of $ I_0 $, can be written by relation
\begin{equation}
 \label{eq:Iobs}
	\mathrm I_{obs}=I_{0}\,g^3\,.
   \end{equation}
Therefore, the locally measured spectral radiance of the CMB at $\nu_ {\mathrm{obs}} $ can be written as
\begin{equation}
\label{eq:Ibbobs2}
  \mathrm I_{obs,\nu_{obs}}= \frac{2\,h\,\nu_{obs}^3}{c^2}\frac{1}{e^{\frac{h\,\nu_{obs}}{k\, T_{obs}}}-1}\,,
   \end{equation}
where $h $ is Planck constant, $k$ is the Boltzmann constant and
\begin{equation}
\label{eq:Ttrans}
T_{obs}=g\,T_{cmb}\,.
\end{equation}
It is easy to see that the transformed spectrum retains the character of the blackbody radiation, but now with the temperature $ T_{\mathrm{obs}} $.
From the relation (\ref{eq:SBL}), we can express the flux density of the blackbody radiation from the elementary spatial angle $ \mathrm{d} \Omega$ in the form
\begin{equation}
 \label{eq:tok1}
  \mathrm{d} \Phi = \sigma\,\frac{\mathrm{d}\Omega}{\pi}\,T^4
   \end{equation}
Then, the flux density of CMB coming from the whole local sky of the exoplanet can be expressed by the formula
\begin{equation}
 \label{eq:tok3}
  \mathrm \Phi= \frac{T_{cmb}^4 \,\sigma}{\pi} \int^{4\pi}_{0} g^4\left(\chi, \psi \right)\, \mathrm{d}\Omega\,,
   \end{equation}
where the nonconstant frequency shift factor $g(\chi, \psi) $ is written as a function of polar and azimuthal angular coordinates in the local sky $ \chi $ and $ \psi $, respectively. The incoming CMB transformed by a coordinate-dependent frequency shift will necessarily have a multiblackbody character.

\subsection{Frequency shift of CMB on the exoplanetary local sky}

We consider an exoplanet orbiting a rotating black hole at the Keplerian circular orbit with a radius $ R_{\mathrm{orb}} $. The line element in the Kerr metric describing the spacetime geometry around a stable rotating black hole can be written using Boyer–Lindquist coordinates in the form
\begin{equation}
\label{kerr_metric}
 \mathrm{d}s^2 = -\left(1-\frac{2  r}{\Sigma}\right) \,\mathrm{d}t^2 
  - \frac{4 r a }{\Sigma} \sin^2\theta\,\mathrm{d}t\, \mathrm{d}\phi 
  + \frac{\Sigma}{\Delta}\, \mathrm{d}r^2 
+ \Sigma \,\mathrm{d}\theta^2
  + \left(r^2 + a^2 +\frac{2 r a^2 \sin^2\theta}{\Sigma}\right)\sin^2\theta\, \mathrm{d}\phi^2, 
        \end{equation}
where $ a $ is a dimensionless angular momentum-black hole spin, $\Sigma \equiv r^{2} + a^{2}\cos^{2}\theta$ and $\Delta \equiv r^{2} - 2r + a^{2}$. Here and below, we use the geometric units ($ G = c = M = 1 $) unless explicitly stated otherwise.
The radial coordinate of the astrophysically relevant outer event horizon of a Kerr black hole is given by the relation
\begin{equation}
R_{\rm +}=1+\sqrt{1-a^2}\,.
\end{equation}

The radial coordinate of the ISCO in the Kerr spacetime is given by
\begin{equation}
\label{RISCO}
R_{\rm ISCO\, \pm}=(3+Z_2\mp\sqrt{(3-Z_1)(3+Z_1+2Z_2)})\,,
\end{equation}
where $Z_1=1+(1-a^2)^{1/3}((1+a)^{1/3}+(1-a)^{1/3})$, $Z_2=(3a^2+Z_1^2)^{1/2}$ and the subscripts $+$ and $-$ denote the corotating and counterrotating case, respectively \citep{Bardeen1972}.
For our further consideration, the ISCO position is crucial, because depending on the black hole spin, it limits the closest approach of the circularly orbiting exoplanet to the outer event horizon of the black hole. In the Schwarzchild nonrotating case with zero spin, the ISCO is located at $R_{\rm ISCO}(a=0)=6$. While the radial coordinate of the corotating ISCO approaches the outer event horizon with increasing spin $a$, and, in the case of extreme spin $ a = 1 $, it merges with the horizon in $R_{\rm +}(a=1)=1$, the counter-rotating ISCO is shifted away with increasing spin. As shown by \cite{BlackSun} and discussed in the Section 1, the heating of the black hole exoplanet orbiting at Schwarzchild ISCO ($R_{\rm ISCO}(a=0)=6$) caused by the amplified CMB is negligible. The CMB-induced heating starts to be effective only at the orbits in close vicinity of corotating ISCOs of rapidly spinning black holes, and therefore at the orbits located very deep in the black hole gravitational field. In such orbits, the role of special relativistic Doppler effect in CMB amplification is also very significant due to very high orbital velocity (see, e.g., \cite{Bardeen1972}).  On the contrary, at the lowest counter-rotating circular stable orbits with the radial coordinate $R_{\rm ISCO\,-}(a>0)>6$, which emerge from the gravitational field with increasing spin, the heating effect will inevitably be weakening with increasing spin. Let us also add that at $R_{\rm ISCO\,-}$ the orbital velocity decreases together with increasing spin \citep{Bardeen1972} and therefore the influence of special relativistic Doppler effect onto CMB amplification also weakens. This is a very natural reason why we have restricted our considerations to orbits corotating with the black hole spin in our further analysis.

Let us introduce a local frame comoving with the exoplanet and equipped with a orthonormal basis with the standard orientation, i.e. with space axes $\langle r \rangle$, $\langle \theta \rangle$, $\langle \phi \rangle$ oriented in the direction of the maximum change of the appropriate Boyer–Lindquist coordinate. 
In the local sky of the exoplanet, the source of the incoming curved ray is projected in the direction of the locally measured photon four-momentum $ p^{\langle \mu \rangle} $ tangent to the ray. Using the unit normalization of the locally measured photon energy, $E_{obs} = - p_{\langle t \rangle} = 1$ , we can write the local covariant components of the four-momentum of the photons, whose source is projected to the local sky with the local polar coordinate $ \chi $ and with the local azimuthal coordinate $ \psi $, in the form
\begin{equation}
 \label{localfourmomentum}
  p_{\langle t \rangle} = -1\,,\qquad
  p_{\langle \phi \rangle} = -\sin{\chi}\cos{\psi}\,,\qquad
  p_{\langle \theta \rangle} = -\cos{\chi}\,, \qquad
  p_{\langle r \rangle} = -\sin{\chi}\sin{\psi}\,.
 \end{equation}
The transformation into the above defined Keplerian local frame can be obtained by a composition of the known transformation into the Zero Angular Momentum Observer (ZAMO) frame and the Lorentz boost by the velocity at a Keplerian corotating circular orbit measured with respect to ZAMO. Recall that Zero Angular Momentum Observers (ZAMOs) conserve their radial coordinate, but they are azimuthally dragged by the rotation of spacetime with the angular velocity $\Omega_{\rm ZAMO}=-g_{\rm t\phi}/g_{\rm \phi\phi}$ measured relative to static observers at infinity. The tetrad of 1-forms of the ZAMO frame in the equatorial plane $ (\theta = \pi / 2) $ can be written in the form \citep{Bardeen1972}
\begin{equation}\label{ZAMO}
	\omega^{(t)}=\left\{\sqrt{\frac{\Delta\Sigma}{A}},0,0,0 \right\},\,\,
	\omega^{(r)}=\left\{0,\sqrt{\Sigma/\Delta},0,0 \right\},\,\,  
	\omega^{(\theta)}=\left\{0,0,\sqrt{\Sigma},0 \right\},\,\,
	\omega^{(\varphi)}=\left\{-\frac{2ar}{\sqrt{A \Sigma}},0,0,\sqrt{\frac{A}{\Sigma}}\right\},     
\end{equation}
where $A \equiv (r^2 + a^2)^2 - a^2\Delta$. The Keplerian orbital velocity, $\beta$, measured by ZAMO located at the radial coordinate of the corotating orbit circular reads \cite{Bardeen1972}
\begin{equation}
\label{eq:beta}
 \beta=\frac{r^2+a^2-2a\sqrt{r}}{\sqrt{\Delta}(r^{3/2}+a)}.
  \end{equation}
Then, the tetrad of 1-forms of the frame associated with the exoplanet at a Keplerian corotating circular orbit can be obtained by the relation 
\begin{equation}
\label{eq:trans_omega}
\omega^{\langle \mu \rangle}_{\,\,\,\,\,\,\,\nu}= \Lambda^{\langle \mu \rangle}_{\,\,\,\,\,\,\,(\alpha)}\omega^{(\alpha)}_{\,\,\,\,\,\,\,\nu}  \, 
\end{equation}
where $\Lambda^{\langle \mu \rangle}_{\,\,\,\,\,\,\,(\alpha)}$ is a matrix of the Lorentz boost in the direction of the azimuthal coordinate.
The resulting tetrad takes the form
\begin{equation}
 \label{eq:ZAMO&Lboost}
	 \omega^{\langle t \rangle}=\gamma \left\{\omega_{\,t}^{(t)}-\beta\omega^{(\phi)}_{\,t},0,0,-\beta\omega^{(\phi)}_{\,\phi} \right\},\,\, 
	  \omega^{\langle r \rangle}=\omega^{(r)}, \,\,   
		\omega^{\langle\theta\rangle}=\omega^{(\theta)},\,\,  
	   \omega^{\langle\phi\rangle}=\gamma \left\{\omega^{(\phi)}_{\,t}-\beta\omega^{(t)}_{\,t},0,0,\omega^{(\phi)}_{\,\phi} \right\}\,,   
       \end{equation}
where $\gamma \equiv (1-\beta^2)^{-1/2}$.
In the Kerr asymptotically flat spacetime, the conserved covariant time component of the photon four-momentum is equal to the negatively taken photon energy at infinity.
The transformation of covariant components of four-momentum can be written by relation
\begin{equation}
\label{finalTrans}
p_{\nu}=\omega^{\langle \mu \rangle}_{\,\,\,\,\,\,\,\nu}\, p_{\left\langle \mu\right\rangle}\,.
\end{equation}
Then, we can easily express the frequency shift of CMB photons coming onto the local sky from a distant universe by the relation
\begin{equation}
g=\frac{p_{\langle t \rangle}}{p_t}=\frac{1}{\omega^{\langle \mu \rangle}_{\,\,\,\,\,\,\,t}\, p_{\left\langle \mu\right\rangle}}\,.
\end{equation}
By substituting the relations (\ref{localfourmomentum}), we obtain the explicit formula for the CMB frequency shift factor as a function of angular coordinates on the local sky in the form
\begin{equation}
 \label{eq:gfactorfinal}
  g\left(\chi, \psi\right)=\frac{\sqrt{1-\beta^2}}{X-Y\sin{\chi}\cos{\psi}}
    \end{equation}
where coefficients
\begin{equation}
X=\omega^{(t)}_{t}-\beta\omega^{(\phi)}_{t},\qquad
Y=\omega^{(\phi)}_{t}-\beta\omega^{(t)}_{t}
\end{equation}
and velocity $ \beta $ must be naturally evaluated at the radial coordinate of the exoplanet orbit ($r=R_ {orb} $) and in the equatorial plane ($ \theta = \pi / 2 $) in which the Keplerian corotating circular orbit lies.

\section{Numerical methods}

The flux integral (\ref{eq:tok3}) was evaluated numerically. The local sky was mapped to a rectangular grid of pixels with a resolution of $ 2n\, x\, n $ using the Mollweide projection. Let us keep in mind that the Mollweide projection transforms the angular coordinates in the sky into a pair of Cartesian coordinates $ x $, $ y $ in a plane in which the sky is projected into an ellipse with the length of the main half axis twice as long as the minor half axis. Considering the unit main half axis, the transformation between the Cartesian coordinates of the Mollweide sky projection and the angular coordinates of the local sky can be written in a simple form:
\begin{equation}
\label{Moltrans}
\chi =\pi/2-\arcsin {\frac {2\xi +\sin 2\xi }{\pi }}\,,\qquad
\psi =\frac {\pi x}{\cos \xi }\,,
\end{equation}
where $\xi =\arcsin {2\,y}$ \citep{MollweideProjection}.
The number of pixels $ N $ contained within the ellipse representing the local sky of the exoplanet reads
\begin{equation}			
N=\frac{1}{2}\pi\,n^{2}\,.
\end{equation}
As the Mollweide projection conserves the area, we can express the solid angle subtended by one grid pixel $ \Omega_{\mathrm{pix}} $ as
\begin{equation}			
\Omega_{pix}=\frac{4\pi}{N}=\frac{8}{n^{2}}
\end{equation}
Then, we can rewrite the integral (\ref{eq:tok3}) as the sum of the contributions of the individual grid pixels in the form
\begin{equation}
\label{eq:Sum}
\Phi=\frac{8\,\sigma\,T^{4}_{cmb}}{\pi\,n^{2}} \sum_{i=1}^n \sum_{j=1}^{2n}  g^{4}_{ij}(\chi_{ij},\psi_{ij})\,,
\end{equation}
where $ i $ and $ j $ is the row and column grid indices, respectively. Naturally, pixels outside the ellipse of the local sky projection and also the pixels corresponding to the black hole shadow (See the Appendix for details) should be excluded from the sum calculation.
 
\begin{figure}[h]  
\begin{center}
\includegraphics[width=1.00\linewidth, angle=0]{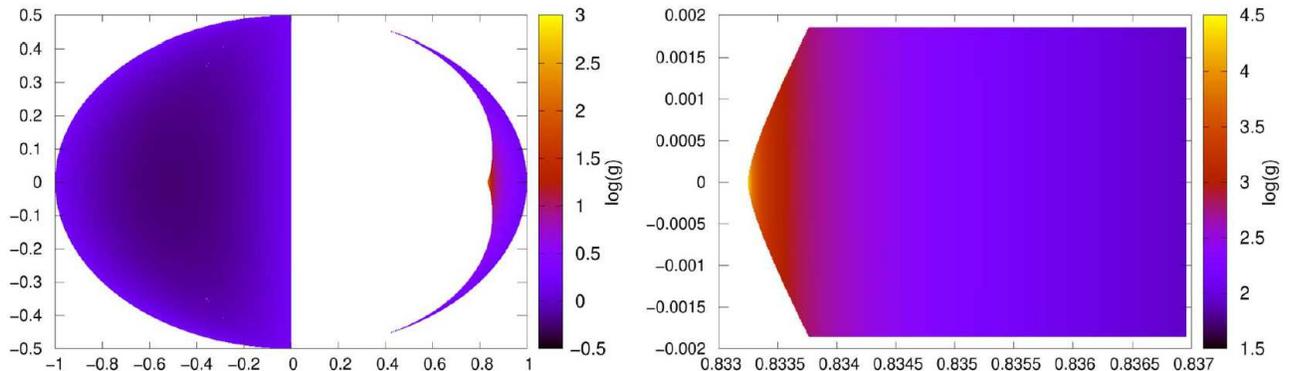}
\end{center}
\caption{Map of the CMB frequency shift factor in the local sky of the exoplanets orbiting a Kerr black hole on the hot edge of the habitable zone ($ R_{orb} = 1.00042$) in the Molweide projection (See Eg. (\ref{Moltrans})). The false colors correspond to the decadic logarithm of the CMB frequency shift. Top: The first iteration: the grid covers the entire sky. The shadow of a black hole fills a significant part of the sky. Bottom: final zoom to calculate the CMB flux density. The detected maximum of the CMB frequency shift is more than of an order higher than in the case of the first iteration.} 
\label{fig:gmap}
\end{figure}

In our numerical modeling, we assume, in accordance with the results of \cite{BlackSun}, that in the case of the black hole exoplanet significantly heated by CMB, a predominant part of the CMB energy flux comes from an extremely small area of the local sky situated on the right edge of the black hole shadow on the equator. To obtain a relevant result of numerical flux integration, it is necessary to zoom the grid onto a region of the local sky so small that the grid resolution is sufficient to identify a small sky region with extreme values of the frequency shift factor. However, the zoomed area must not be too small to neglect the CMB flux from the vicinity of the maximum emitting pixels, which also contributes significantly to the total flux (See Fig \ref{fig:gmap}). During each iteration, the zoomed area of the local sky is always centered relative to the point in the local sky where the CMB frequency shift factor reaches its strong maximum. Its coordinates are determined numerically by searching for the maximum of the function (\ref{eq:gfactorfinal}).

The optimum zoom of the grid providing the maximum value of the sum (\ref{eq:Sum}) is set in the code by a series of iterations by using Brent’s method for maximization \citep{NumericalRecipes}. Let us note that reducing the horizontal range of the grid relative to the major ellipse half axis by factor $k$ also means reducing the magnitude of the solid angle subtended by one pixel of the grid by a factor of $k^2$. Therefore, the sum (\ref{eq:Sum}) must be divided by the corresponding $k^2 $ factor at each iteration. The used method converges with increasing grid resolution with respect to the value of the sum (\ref{eq:Sum}) and thus also with respect to the value of the temperature (\ref{eq:exotemp}). A resolution setting of $ n =  500$ was used, further increasing the number of pixels of the grid already makes the estimation of the exoplanet temperature more precise only in the order of tenths of degrees.

\section{Habitable zone}

\begin{figure*}[h!]  
\begin{center}
\includegraphics[width=1.0\linewidth]{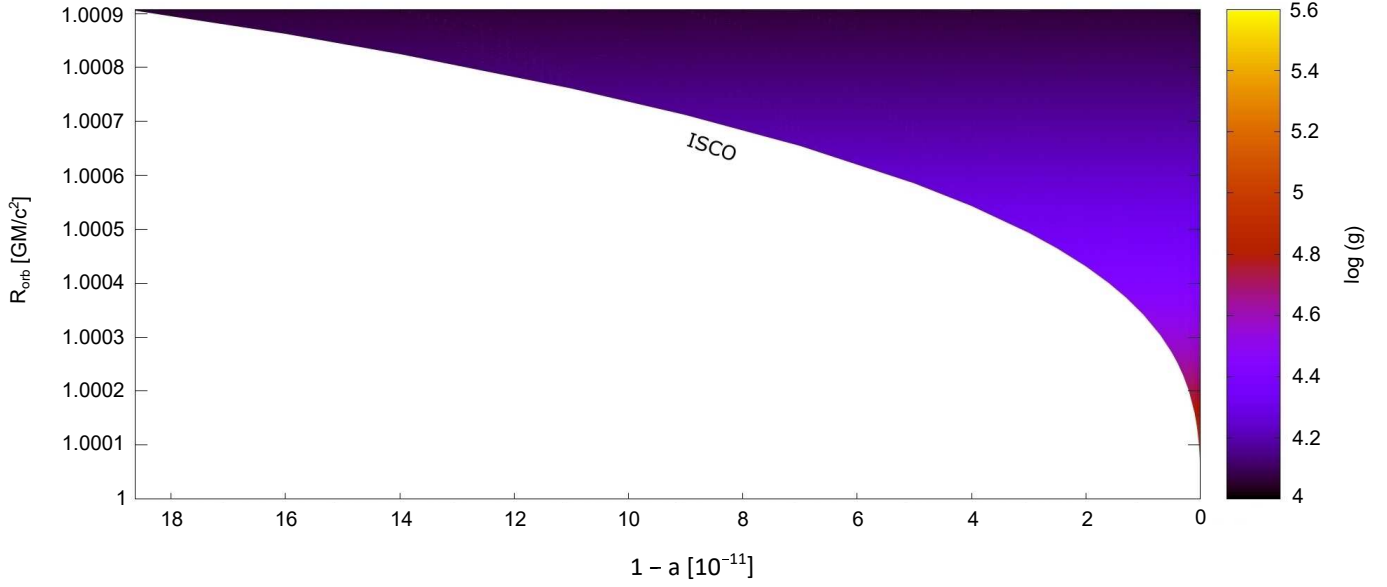}
\end{center}
\caption{Maximum frequency shift factor of the CMB coming from the exoplanetary local sky $g$ (in a decadic logarithmic scale) as a function of the black hole spin $a$ and the radius of the exoplanetary orbit $R_{\mathrm{orb}}$.}
\label{fig:result_g}
\end{figure*}

\begin{figure*}[h!]  
\begin{center}
\includegraphics[width=1.0\linewidth]{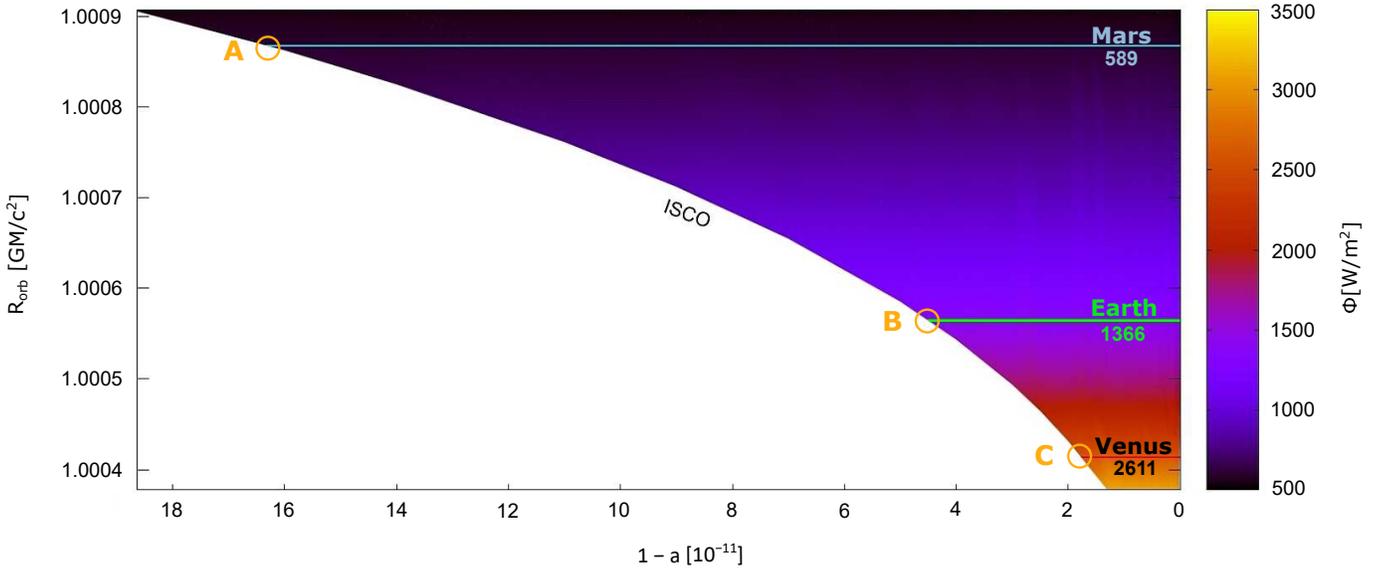}
\end{center}
\caption{Flux density of the CMB coming from the exoplanetary local sky (solar constant) $ \Phi $ [$ W / m^{2} $] as a function of the black hole spin $a$ and the radius of the exoplanetary orbit $R_{orb}$. The range of $R_{\mathrm{orb}}$ is restricted to the habitable zone only.}
\label{fig:result_SC}
\end{figure*}
\begin{figure*}[h!]  
\begin{center}
\includegraphics[width=1.0\linewidth]{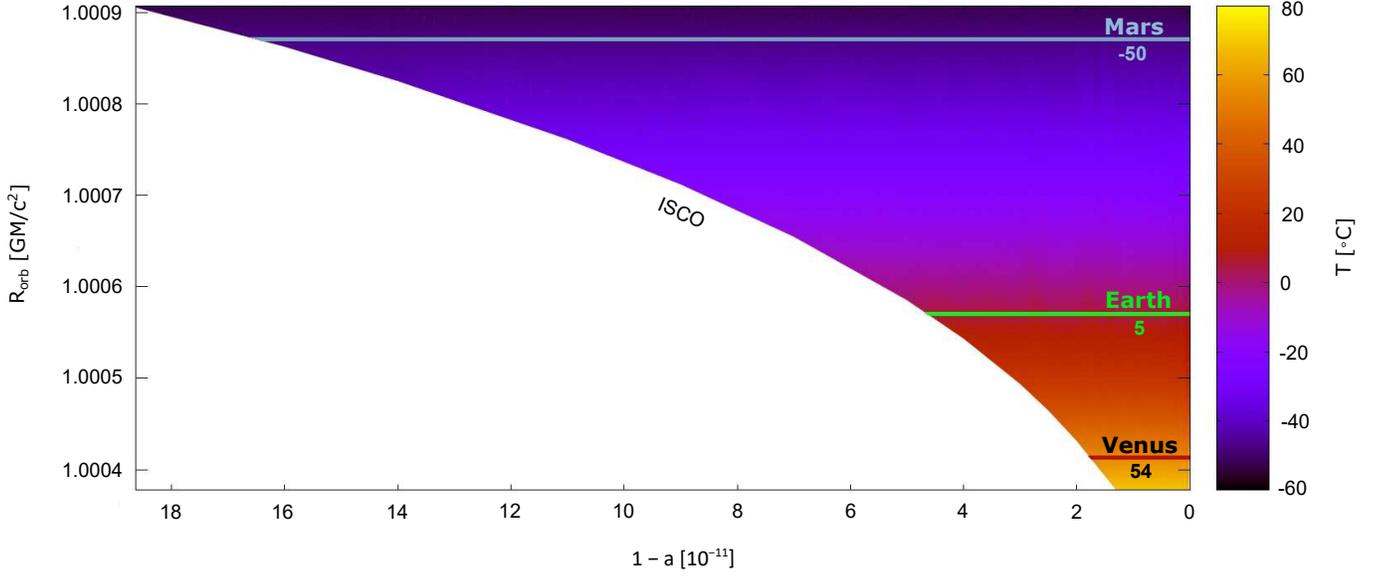}
\end{center}
\caption{Estimated temperature of the exoplanet $T$ [$^\circ C$] as a function of the black hole spin $a$ and the radius of the exoplanetary orbit $R_{\mathrm{orb}}$.The range of $R_{\mathrm{orb}}$ is restricted to the habitable zone only.}
\label{fig:result_T}
\end{figure*}
\begin{figure*}[h!]  
\begin{center}
\includegraphics[width=1.0\linewidth]{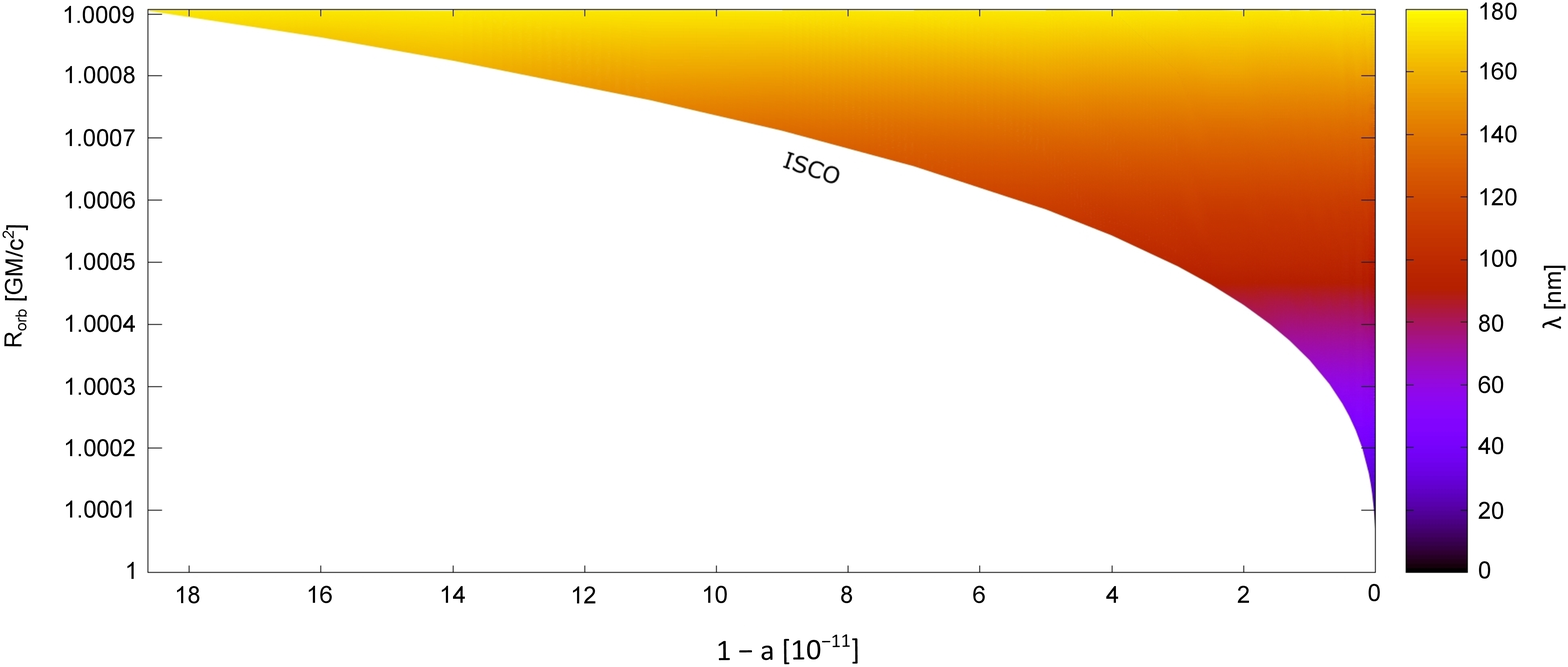}
\end{center}
\caption{
Peak wavelength of the resulting multiblackbody spectrum of incoming CMB $\lambda_{max}$ [$nm$] as a function of the black hole spin $a$ and the radius of the exoplanetary orbit $R_{\mathrm{orb}}$.}
\label{fig:result_lambda}
\end{figure*}

Using the numerical methods described in the previous section and implemented in the LSDPlus software package designed for modeling optical effects in strong gravity \citep{twinpeaks}, we mapped the CMB flux density as a function of a black hole spin and radial coordinates of the Keplerian circular orbit of the exoplanet. Along the CMB flux density, the estimated temperature of the exoplanet (\ref{eq:exotemp}), the maximum CMB frequency shift factor, and the profile of the resulting multiblackbody CMB energy spectrum were also calculated.
 
For the existence of life based on known biology, the habitable zone is often considered to allow the presence of water in the liquid state. Such a habitable zone is, in a very rough approximation, bounded by conditions at the orbit of Mars where the mean value of the solar constant is $\Phi\approx589 W/m^{2}$ and by the conditions at the orbit of Venus where the mean value of the solar constant is $\Phi\approx2611\,  W/m^{2}$. At the orbit of the Earth, which lies at the center of the habitable zone defined in this way, the mean value of the solar constant reaches $\Phi\approx1366\, W/m^{2}$ \citep{HabitableZone}. The temperature of Mars, Earth, and Venus estimated by the relation (\ref{eq:exotemp}) is about $ -50\,^{\circ} $C, $ 5\, ^{\circ}$C and $54\,^{\circ}$C, respectively.

The results of numerical modeling show that the heating of black hole exoplanets corresponding to the conditions in the above defined habitable zone can only be achieved for a very high black hole spin and at circular orbits located very close to ISCO. Moreover, the results clearly show that the CMB energy flux density within habitable zone depends only on the radial coordinate of the exoplanetary circular orbit. The results of numerical modeling are illustrated in the Figs. \ref{fig:result_g} - \ref{fig:result_lambda}. The maps of the CMB energy flux density (see Fig. \ref{fig:result_SC}) and the temperatures of the exoplanet (see Fig. \ref{fig:result_T}) are restricted to the habitable zone defined above. Its boundaries corresponding to the values of the appropriate quantities at the orbit of Mars and Venus are marked by the blue and red contour lines, respectively. The green contour line within the habitable zone corresponds to the conditions at the orbit of the Earth. 

For a black hole with the spin $ a =1-1.64 \cdot 10^{-10}$, the energy flux density of in the coming CMB is equal to the mean value of the solar constant at the Mars orbit for the case of the exoplanet orbiting just at ISCO with $ R_{\mathrm{orb}} = 1.00090 $. This spin value thus determines the lower limit of the range of possible spin values allowing the existence of a black hole habitable zone (See point A in Fig. \ref{fig:result_SC}). 
Moreover, such a value of the radial coordinate delimites the cold edge of the habitable zone. The maximum CMB frequency shift reaches the value of $ g = 11,450 $ here.
The range of possible radii of exoplanetary orbit in the habitable zone increases with growing spin. For a black hole with the spin $ a = 1-4.5 \cdot 10^{-11}$, the energy flux density of the incoming CMB is equal to the mean value of the solar constant at the Earth's orbit for the case of the exoplanet orbiting just at ISCO with $ R_{\mathrm{orb}} = 1.00057 $ (See point B in Fig. \ref{fig:result_SC}). The maximum CMB frequency shift reaches the value of $ g = 18,260 $ here. 
Finally, for a black hole with spin $ a = 1-1.75 \cdot 10^{-11}$, the energy flux density of the incoming CMB is equal to the mean value of the solar constant at the orbit of Venus for the case of the exoplanet orbiting just at ISCO with $ R_{\mathrm{orb}} = 1.00042 $ (See point C  in Fig. \ref{fig:result_SC}). Therefore, such a value of the radial coordinate delimits the hot edge of the habitable zone. The maximum CMB frequency shift reaches the value of $ g = 24,742 $ here.

\section{Resistance to tidal disruption}
Exoplanetary circular orbits in the habitable zone are located very close to the event horizon. 
Evidently, we can assume that a black hole must be supermassive enough, so that a reasonably large exoplanet would at all fit into a low orbit and would not touch the event horizon. 
However, the mass of the black hole is also related to more subtle requirement for the magnitude of tidal forces that should not disrupt the exoplanet.
The radius $R_t$, at which the tidal disruption of a smaller body of mass, $m_{\star} $, and the radius, $r_{\star}$, orbiting around a black hole of the mass, $M$, will occur, can be roughly approximated by the relation \citep{1988Natur.333..523R}
\begin{equation} \label{tidalradius}
R_t= \left(\frac{M}{m_{\star}}\right)^{\frac{1}{3}}\,r_{\star}\,.
\end{equation}
Considering an exoplanet of Earth's mass and radius ($m_{\star}=5.97\cdot 10^{24}$kg , $r_{\star}=6.37 \cdot 10^6$ m) orbiting at the hot edge of the habitable zone, the minimum mass of the black hole at which the exoplanet will be resistant to tidal forces is $M \sim 1.63 \cdot 10^8\,m_{\odot}$. Throughout the habitable zone, this value changes quite negligibly. Therefore, we need a supermassive black hole (at least) $\sim 37$ times more massive than Sagittarius A* in the Milky Way center with the mass estimation $\sim 4.5 \cdot 10^6\,m_{\odot}$ \citep{2008ApJ...689.1044G}. Then, in the case of an extreme Kerr black hole ($ a = 1 $), the distance of the hot edge of the habitable zone from the event horizon will be $1.022 \cdot 10^5\,km$, and the habitable zone width will be $1.667 \cdot 10^5\,km$. The corresponding radius of outer event horizon is $ 1.613 \, au$.

\section{Spectral properties of incoming CMB}

\begin{figure}[h!]  
\begin{center}
\includegraphics[angle=0,width=0.9\linewidth]{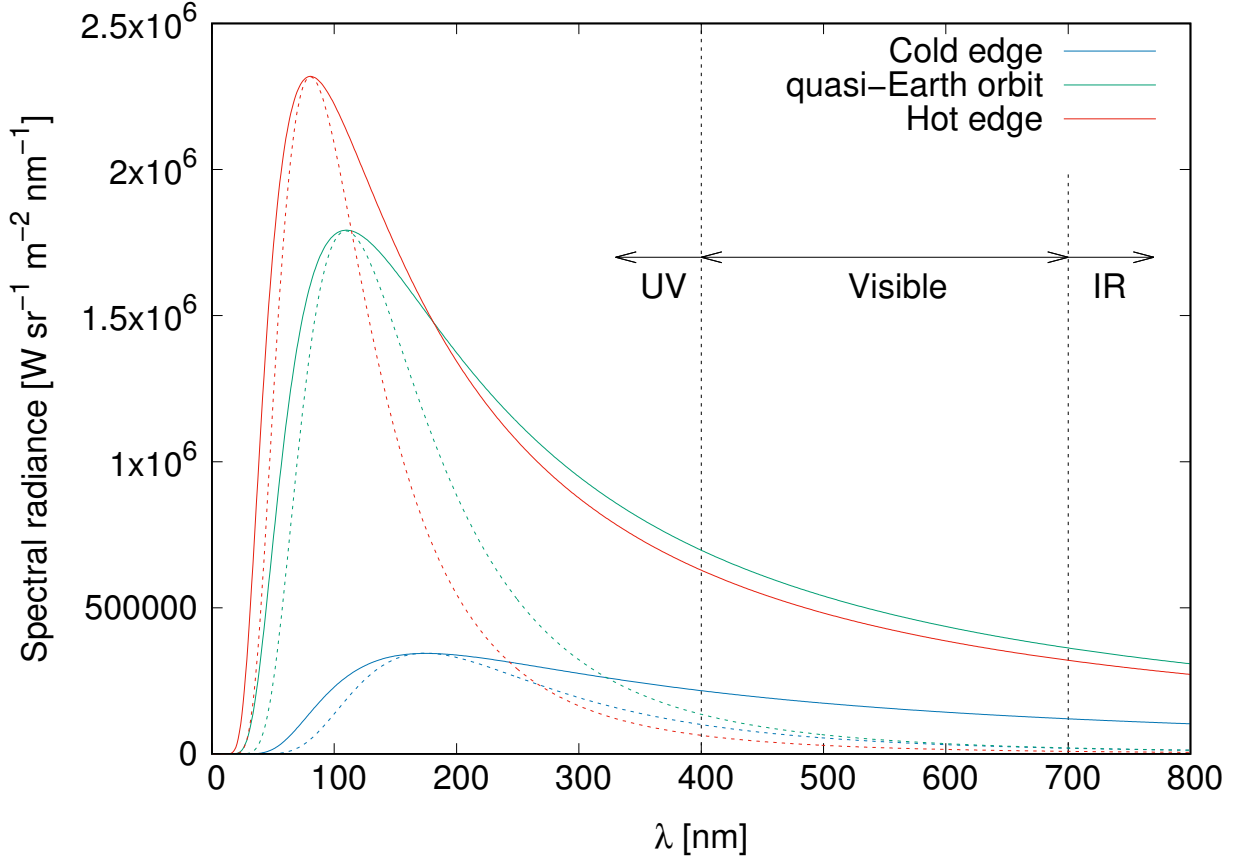}
\end{center}
\caption{Multiblackbody spectral profiles of CMB coming from the local sky of the exoplanet orbiting at the cold edge of the habitable zone (blue, $\lambda_{max}=175\,nm$), at the orbit corresponding to Earth's thermodynamic temperature, (green, $\lambda_{max}=110\,nm$) and at the hot edge (red, $\lambda_{max}=80\,nm$) of the habitable zone. Corresponding normalized blackbody radiation spectra with the same wavelength of the spectral peak $ \lambda_ {max}$ are plotted by dashed curves.
}
\label{fig:spectra}
\end{figure}

As illustrated in Fig. \ref{fig:result_lambda}, the map of the spectral peak wavelength $\lambda_{max} $ as a function of the spin and the orbital radius, shows that the peak of the resulting multiblackbody energy spectrum of the incoming CMB is always located in the ultraviolet band ($\lambda_{max}\in(80,175)\,nm$) for the case of the black hole exoplanets orbiting in the habitable zone.
The whole profile of the resulting multiblackbody spectrum is illustrated in Fig. \ref{fig:spectra} for three distinctive cases. The solid curves in the figure correspond to the multiblackbody spectra of the incoming CMB, the dashed curves correspond to the pure blackbody spectra with the spectral peak located also at the $\lambda_ {max}$ and normalized to the maximum of the CMB spectral radiance. Blue and red profiles correspond to the cool and hot edge of the habitable zone, respectively.  The appropriate spectral peaks are located at $ \lambda_{max} = 175 \, nm $ and $\lambda_ {max} = 80 \, nm $. Green profiles correspond to the orbit of an exoplanet with a thermodynamic temperature equal to Earth\'{ }s. In this orbit, the spectral peak is located at $ \lambda_{max} = 110 \, nm $.  It is clearly visible that the spectral peaks of the multiblackbody spectra CMB are considerably wider than the peaks of the corresponding pure blackbody spectral profile in all the illustrated cases. Therefore, a significant part of the flux comes in the visible and infrared bands.

\section{Time scales}

\subsection{Time dilation}

\begin{figure}[h]  
\begin{center}
\includegraphics[width=0.7\linewidth, angle=0]{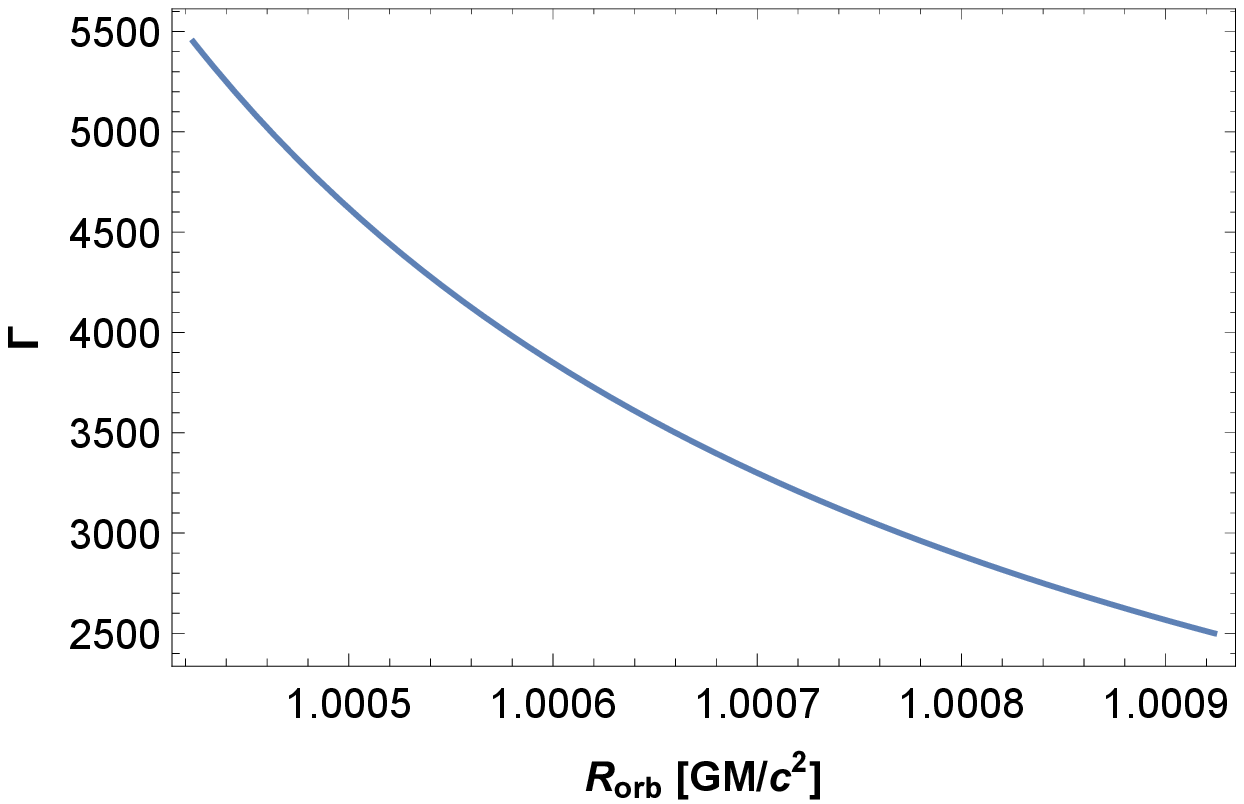}
\end{center}
\caption{Time dilation factor $\Gamma$ as a function of the radial coordinate of the exoplanetary circular orbit. The range of the radial coordinate si restricted to the habitable zone, the spin is fixed to the extreme value of $a=1$.} 
\label{fig:dilation}
\end{figure}

The time dilation in the frame associated with an exoplanet at the Keplerian circular orbit can be understood as the combination of the time dilation in the ZAMO frame determined by inverse of the ZAMO 1-forms component $\omega^{(t)}_{\,\,\,\,\,\,\,t}$ and the special relativistic time dilation due to the Lorentz boost by the orbital velocity $\beta$ (\ref{eq:beta}). Therefore, the time dilation factor $\Gamma$ can be explicitly expressed by the formula 
\begin{equation}\label{TDF}
	\Gamma=\gamma\sqrt{\frac{A}{\Delta\Sigma}}\,.
\end{equation}
In the habitable zone, the dependency of the time dilation factor on the spin is negligible. On the contrary, the dependency of the exoplanetary orbit on the radial coordinate is significant, as illustrated in the Fig. \ref{fig:dilation}. The time dilation factor reaches the value of $\Gamma \sim 2500$ on the cold edge of the habitable zone and it grows up to the value of $\Gamma \sim 5450$ on the hot edge. 

\subsection{Time evolution of CMB}

The contemporary cosmological models predict that the CMB decreases its temperature by one degree per $\sim 6.35 \cdot 10^9$ yr (see, e.g., \cite{2010arXiv1005.0555S} for details). This implies that the cold edge of the habitable zone will be located at the current radial coordinate of the hot edge in $\sim 5.6 \cdot 10^9$ yr, when the CMB cools down to $1.87$ K. At the same time, the hot edge will be shifted in the very close vicinity of the event horizon. Considering the time dilatation effect discussed above, the exoplanet orbiting at the stable circular orbit near the hot edge leaves the habitable zone in $\sim 10^6$ yr of local time, which certainly limits the perspectives of life evolution.

\subsection{Orbital decay by gravitational radiation}

Assuming the mass of the black hole ensuring the resistance to tidal disruption of the Earth-sized exoplanet, the value of the mass ratio governing a magnitude of the gravitational emission is $5.4 \cdot 10^{-13}$. Extrapolating characteristic time scale of the orbital decay by gravitational radiation in the Kerr field numerically modeled for the case of the test particle with the mass ratio $10^{-2}$ \citep{2018EPJWC.16802006R}, we found that the the orbital decay of the exoplanet is very slow. The roughly estimated time scale of the gravitational orbital drift through the habitable zone is in order of $\sim 10^{10}$ yr. The similar scale is also estimated for the time to the impact on the event horizon for the exoplanet orbiting on the hot edge of the habitable zone.

\section{Conclusion}

A habitable zone, where the water in the liquid state can exists on the surfaces of hypothetical Earth-like exoplanets, can arise in the close vicinity of the event horizon of spinning supermassive black holes with $M \geq 1.63 \cdot 10^8\,m_{\odot}$ and a very high spin $a \geq 1-1.64 \cdot 10^{-10}$. The cold and hot edge of such habitable zone correspond to a circular Keplerian orbit with radius $R_{orb}=1.00090\,GM/c^{2}$ and $R_{orb}=1.00042\,GM/c^{2}$, respectively. Naturally, the possibility of life existence in such an extreme environment is associated with many nontrivial questions.

The very existence of a sufficiently supermassive black hole with almost extreme spin and the subsequent formation of an exoplanet near the event horizon is questionable. Moreover, a collision with another orbiting or accreted matter can be certainly relevant for the threat to life existence on the surface of an exoplanet orbiting deep in the strong black hole gravitational field. Therefore, a supermassive black hole successfully hosting a habitable exoplanet should be old enough to accrete all the surrounding cosmic garbage. It certainly could not be even an occasionally active galactic nucleus. Even an exoplanet collision with a meteor or even an asteroid moving in an extreme gravitational field with relativistic velocity would probably be fatal to life on an black hole exoplanet. We note, that the absence of counterrotating circular orbits in such a close vicinity of the event horizon makes the situation somewhat more optimistic.

The resulting multiblackbody spectral profile of the incoming CMB  with the peak in the ultraviolet band can be limiting for the evolution of biological life as we know it on Earth. The evolution of life on a black hole exoplanet orbiting deep in the extreme gravitational field would also be limited by the shortening of relevant time scales caused by the relativistic time dilation which factor reaches the value of thousands. 

In this paper, we roughly estimate the black hole mass ensuring the resistance to tidal disruption of the exoplanet. The subsequent study can be devoted to more precise analysis of tidal forces at the Keplerian orbit around a rapidly spinning black hole, whose fully relativistic description is not quite trivial (see, e.g., \cite{2005PhRvD..71d4017I}). In such a study,  the conditions for the tidal locking in the extreme Kerr field also can be analysed. 

We consider heating the black hole exoplanets only by the CMB. In the following studies, the results can be straightforwardly refined by taking into account another component of cosmic background in infrared, optical, and X-ray band (CIB, COB, and CXB), although the microwave component is dominant (see, e.g., \cite{2002astro.ph..2430H}). Furthermore, the radiation of other cosmic objects located in the relative vicinity of a supermassive black hole can also be amplified by the extreme gravitational field. As an illustrative example, let us mention S-type stars  orbiting very close to the Milky Way core \citep{2012Sci...338...84M}. Another possible, and probably more promising scenario, is related to binary systems containing a black hole and a main–sequence star. This could probably create a dynamic habitable zone around stellar black holes and probably with a more life-friendly spectral profile of incoming radiation. However, more detailed computer modeling is necessary to analyze these more complex scenarios (see also \cite{2019arXiv191000940S}).

\acknowledgments
We would like to acknowledge the Czech Science Foundation (GA\v{C}R) grant GA\v{C}R 17-16287S and the internal grant of Silesian University SGS/13/2019. We would like to thank Tomáš Opatrný, Lukáš Richterek, Luigi Stella, and Vladimír Karas for inspiring discussions and remarkable notes. We also would like to thank Osservatorio Astronomico di Roma in Monte Porzio Catone and International Space Science Institute in Beijing, where significant a part of this work was done, for their kind hospitality.

\bibliography{BibTeX1}

\appendix
\section{Black hole shadow in the exoplanetary local sky}

The black hole shadow in the local sky of a black hole exoplanet is formed by photon trajectories coming from the very close vicinity of the outer event horizon for which there is no radial turning point between the outer event horizon and the orbital radius of exoplanet $ R_{\mathrm{orb}} $. Let us keep in mind, that the boundary of a rotating black hole shadow is a projection of an area of unstable spherical photon orbits wrapping the black hole. The shape of the shadow of a Kerr black hole is strongly dependent on the spin, the inclination of the observer, and the reference frame associated with the observer. As the observer approaches the outer event horizon, the shadow area in the sky increases significantly (see, e.g., \cite{Chandra1983,Viergutz1993,2003GReGr..35.1909T,BlackSun} for details).

The components of the photon four-momentum in the Kerr geometry reads \citep{Carter} 
\begin{align}
     p^{r} &= \dot{r} = s_r\Sigma^{-1} \sqrt{ \mathcal{R}_{l,q}(r)}\,,\label{CarterEQs}\\
     p^{\theta} &= \dot{\theta} = s_{\theta}\Sigma^{-1} \sqrt{\Theta_{l,q}(\theta)}\,,\nonumber\\
     p^{\phi} &= \dot{\phi} = \Sigma^{-1} \Delta^{-1}
     \left[ 2ar + l\left( \Sigma^{2} - 2r \right)  
\mathrm{cosec}^{2} \theta \right]\,,\nonumber\\
     p^{t} &= \dot{t} = \Sigma^{-1} \Delta^{-1} \left( \Sigma^{2} -
     2ar l \right)\,,\nonumber
\end{align} 
where dots denote the derivative with respect to the affine parameter and the sign pair $s_{r}$,$s_{\theta}$ describes the orientation of the radial and latitudinal motion.
Radial and latitudinal effective potentials are given by the relations
\begin{align} 
       \mathcal{R}_{l,q} \left( r \right) &=  \left( r^{2} + a^{2}
 - a l\right) ^{2}
       - \Delta \left[ q + \left( l - a \right) ^{2} \right]\,,\nonumber \\
       \Theta_{l,q} \left( \theta \right) &= q + a^{2} \cos^{2}
\theta -l^{2} \mathrm{cot}^{2} \theta\,.\label{eqn:2.1.3}
\end{align} 
The constants of motion $l$ and $ q$ are related to the covariant photon four-momentum components by the relations 
\begin{equation} \label{constant_def}
  \begin{split}
l &= -\frac{p_{\phi}}{p_{t}}, \,\\
q &= \left(\frac{p_{\theta}}{p_{t}}\right)^{2} + \left(l\tan\left(\pi/2 - \theta\right)\right)^{2} - a^{2}\cos^{2}\theta\,.    
  \end{split}
\end{equation}
The covariant photon four-momentum components can be obtained from local components (\ref{localfourmomentum}) using the transformation (\ref{finalTrans}), which relates the pair of angular coordinates in the exoplanetary local sky ($ \chi $, $ \psi $) to the corresponding pair of photon motion constants ($l $, $ q $).
Then we can easily identify the range of coordinates $ \chi $ and $ \psi$ in the exoplanetary local sky corresponding to the shadow of the black hole. Photon trajectories that form the shadow of a black hole must have a positive radial sign ($ s_{r} = 1 $) and in the range $r \in ({R_{\rm +},R_{orb}})$ they must fulfill the condition $\mathcal{R}_{l,q}(r)\geq 0$ corresponding precisely to the absence of a radial turning point between the event horizon and the exoplanetary circular orbit.


\end{document}